\documentclass[letterpaper, 10 pt, conference]{ieeeconf}  

\usepackage{nicefrac}
\usepackage{siunitx}
\usepackage{svg}
\usepackage{booktabs}
\usepackage{array,framed}
\usepackage{booktabs}
\usepackage{
  color,
  float,
  epsfig,
  wrapfig,
  graphics,
  graphicx,
  subcaption
}
\usepackage{textcomp,amssymb}
\usepackage{setspace}
\usepackage{latexsym,fancyhdr,url}
\usepackage{enumerate}
\usepackage{algorithm2e}
\usepackage{algpseudocode}
\usepackage{graphics}
\usepackage{xparse} 
\usepackage{xspace}
\usepackage{multirow}
\usepackage{csvsimple}
\usepackage{balance}
\usepackage{amsmath}
\usepackage{amssymb}
\usepackage{bm}

\usepackage{
  tikz,
  pgfplots,
  pgfplotstable
}
\usepackage{hyperref}

\usetikzlibrary{
  shapes.geometric,
  arrows,
  external,
  pgfplots.groupplots,
  matrix
}

\pgfplotsset{compat=1.9}

\usepackage{mathtools,}

\IEEEoverridecommandlockouts                              

\title{\LARGE \bf
COMPASS: Cooperative Multi-Agent Persistent

Monitoring using Spatio-Temporal Attention Network
}

\author{Xingjian Zhang$^{1}$, Yizhuo Wang$^{1}$, Guillaume Sartoretti$^{1}$ 
\thanks{$^{1}$Authors are with the Department of Mechanical Engineering, College of Design and Engineering, National University of Singapore
        {\tt\small xingjian@u.nus.edu, wy98@u.nus.edu, mpegas@nus.edu.sg}}%
}

\begin{document}
\maketitle
\pagestyle{empty}
\begin{abstract}

Persistent monitoring of dynamic targets is essential in real-world applications such as disaster response, environmental sensing, and wildlife conservation, where mobile agents must continuously gather information under uncertainty. We propose COMPASS, a multi-agent reinforcement learning (MARL) framework that enables decentralized agents to persistently monitor multiple moving targets efficiently. We model the environment as a graph, where nodes represent spatial locations and edges capture topological proximity, allowing agents to reason over structured layouts and revisit informative regions as needed. Each agent independently selects actions based on a shared spatio-temporal attention network that we design to integrate historical observations and spatial context. We model target dynamics using Gaussian Processes (GPs), which support principled belief updates and enable uncertainty-aware planning. We train COMPASS using centralized value estimation and decentralized policy execution under an adaptive reward setting. Our extensive experiments demonstrate that COMPASS consistently outperforms strong baselines in uncertainty reduction, target coverage, and coordination efficiency across dynamic multi-target scenarios.

\end{abstract}


\section{Introduction}
\label{sec:intro}

Persistent monitoring requires a team of mobile agents to frequently visit dynamic targets to maintain an accurate, shared belief of their states. This problem is central to applications like environmental sensing and disaster response, which operate under strict resource constraints~\cite{ore2015autonomous,reece2010introduction}. A successful strategy must balance spatial coverage—minimizing revisit times—with temporal tracking, selecting viewpoints that best reduce uncertainty as targets move.

We consider multi-agent monitoring in environments where neither target motion models nor spatial distributions are known beforehand.  Our solution, COMPASS, equips each agent with a light-weight yet expressive decision module that combines Gaussian-Process (GP) belief estimation with a spatio-temporal transformer.  The GP layer fuses new observations into probabilistic predictions and uncertainties for every target, while the transformer first aggregates how those beliefs evolve across a sliding temporal window and then reasons over the graph that discretizes free space.  All computations run locally on board, and agents exchange only compact belief updates so that coordination scales gracefully with team size and does not rely on a global map or a central planner.

\begin{figure}[t]
\vspace{0.1cm}
\setlength{\belowcaptionskip}{-5mm}
  \centering
    \includegraphics[width=.48\textwidth]
    {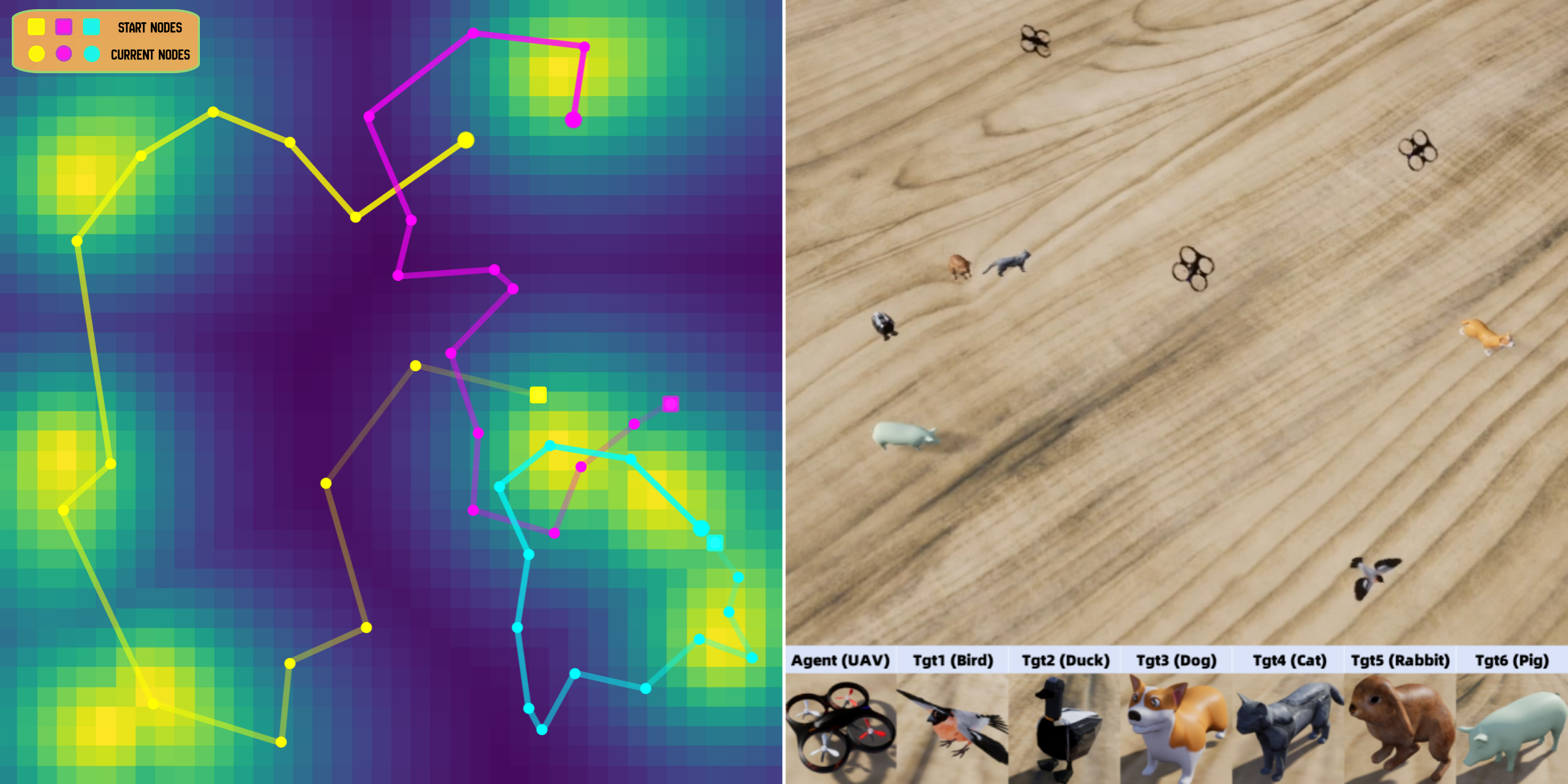}
    \vspace{-0.4cm}
    \caption{
    \textbf{Illustration of the COMPASS framework in a simulated multi-agent wildlife monitoring task.} 
    \emph{(Left)} Visualization of three UAV agents' trajectories and corresponding belief maps over dynamic targets; brighter areas denote higher uncertainty regions that agents are incentivized to explore. 
    \emph{(Right)} AirSim-based 3D deployment, where three UAVs cooperatively track multiple moving animal targets modeled with distinct 3D meshes.}
    \label{illustrate}
    \vspace{-0.1 cm}
\end{figure}

Earlier research illuminates separate facets of this challenge but stops short of a unified solution.  STAMP leverages temporal attention to trade exploration for revisit frequency but assumes a single robot~\cite{stamp}.  MA-G-PPO extends attention mechanisms to multi-agent reinforcement learning (MARL) yet depends on continuous communication and a coarse global occupancy grid~\cite{chen2021multi}.  GALOPP accounts for sensing, localisation, and connectivity constraints in heterogeneous teams, though it lacks a principled way of incorporating historical observations when choosing future actions~\cite{mishra2024multi}. In our work, we address tasks characterized by highly dynamic spatio-temporal dependencies, where both target motions and agent interactions evolve continuously over time. Inspired by recent findings in spatio-temporal reasoning for dynamic robotics and perception tasks~\cite{li2021attention,hu2022spatial}, we design a unified transformer architecture that explicitly encodes temporal evolution and spatial relations in a theoretically grounded manner.   By jointly modelling uncertainty with GPs and coupling spatial and temporal reasoning inside one transformer, COMPASS closes this gap and produces decentralised policies that adapt online to moving targets.

We test COMPASS in a high-fidelity simulator populated with targets that follow partially random trajectories.  Against competitive baselines—STAMP*, Auction, Coverage, and Random—our framework achieves the lowest mean predictive uncertainty, the most balanced visitation statistics, and strong scalability across different graph densities and team sizes, establishing a new reference point for cooperative persistent monitoring.

\section{Related Work}
\label{sec:relwork}

Persistent monitoring problems can be broadly divided into two categories: \textit{field monitoring}, which focuses on estimating and tracking continuous environmental variables over spatial domains, and \textit{target monitoring}, which aims to maintain accurate knowledge of discrete, dynamically moving entities~\cite{popovic2020informative}. Gaussian Processes (GPs) have been widely adopted in field monitoring due to their ability to model spatial-temporal correlations and quantify uncertainty for decision-making~\cite{seeger2004gaussian}. Binney et al.~\cite{binney2010informative} leveraged mutual information to optimize robot trajectories in aquatic environments, while Hitz et al.~\cite{hitz2014fully} applied GPs to enable adaptive online planning in oceanographic sampling tasks. Hollinger and Sukhatme~\cite{hollinger2014sampling} further used GP-based predictions to guide exploration towards regions of high expected information gain.

While early approaches focused on planning optimal visitation sequences to reduce target state uncertainty, many of these are inspired by combinatorial formulations such as variants of the Traveling Salesman Problem (TSP)~\cite{jin2022adaptive}. Smith et al.~\cite{smith2011optimal} introduced uncertainty-aware cost functions to guide robot routing. However, these methods often assume static or predictable target dynamics, limiting their applicability to realistic, dynamic settings~\cite{barfoot2010field}. To address this, adaptive routing methods like the Orienteering Problem (OP) have been explored, which prioritize high-value regions based on changing conditions~\cite{gunawan2016orienteering}. For instance, Yu et al.~\cite{yu2012large} used stochastic TSP variants to tackle uncertain urban air mobility tasks. Despite these advances, such formulations typically focus on one-shot routing and lack the temporal adaptivity required for persistent monitoring of evolving targets.

Reinforcement learning (RL) has emerged as a powerful tool for persistent monitoring, enabling agents to adapt to dynamic environments through trial-and-error learning. Deep RL (DRL) approaches, such as those combining PHD filters with neural policies~\cite{sung2012drift} or incorporating recurrent structures for temporal reasoning~\cite{zhou2024her}, have improved multi-target tracking performance by learning from interaction histories.
Meanwhile, attention mechanisms have shown strong potential for capturing temporal and spatial dependencies in structured domains. Originally proposed for NLP~\cite{vaswani2017attention}, attention has been applied in robotics to improve exploration efficiency~\cite{khan2016cooperative} and planning robustness~\cite{li2021attention}. However, most existing works isolate either spatial~\cite{glock2023spatial} or temporal~\cite{smith2011optimal} reasoning. While some, like Hu et al.~\cite{hu2022spatial}, integrate both in static prediction settings, they are not designed for decentralized multi-agent coordination. These limitations motivate our approach, which unifies spatio-temporal attention within a graph-based reinforcement learning framework for scalable and cooperative target monitoring.

Previous work has highlighted the importance of leveraging spatial and temporal structure for planning under uncertainty~\cite{stamp}. Building on this, our method combines GP-based uncertainty modeling with a Spatio-Temporal Attention Network that encodes both historical temporal dependencies and spatial interactions among agents and targets. Unlike prior approaches, we unify these dimensions into a coherent framework. Additionally, we incorporate agent presence information into the input features, enabling context-rich and adaptive monitoring strategies. This idea is aligned with Miao et al.~\cite{miao2023occdepth}, who demonstrated the benefits of presence-aware planning in multi-agent navigation.


\section{Problem Setup}
\label{sec:background}

We consider the task of multi-agent persistent monitoring, where a team of $M$ autonomous agents monitors $N$ mobile targets $\mathcal{T} = \{T_1, ..., T_N\}$ within a bounded 2D space $\mathcal{W} \subseteq \mathbb{R}^2$. Each target $j$ has an evolving state (e.g. location, field distribution) $x_j(t) \in \mathcal{X}_j$, governed by unknown or stochastic dynamics. Here, $X_j \subseteq \mathbb{R}^2$ denotes the spatial state space of target $j$, representing all possible positions within the monitoring region. The objective is to maintain accurate belief estimates over all target states throughout the mission.

\subsection{Belief Modeling with Gaussian Processes}
\label{subsec:gp}

In this paper, we focus exclusively on the positional dynamics of the targets, i.e., we model only their location changes over time and ignore other potential state variables such as velocity or heading. To model spatio-temporal uncertainty over target locations, we adopt Gaussian Processes (GPs)~\cite{stamp,vasudevan2009gaussian}, maintaining an independent GP $GP_j$ for each target $j$.
The inputs to each Gaussian Process are spatio-temporal vectors 
\[
  \textbf{x} = [\,\mathbf{p},\,\mathbf{t}\,] 
  = [\,p_1, t_1, \dots, p_k, t_k\,],
\]
when aggregating a sector or sequence of $k$ measurements,
while the outputs represent information about target presence: its posterior mean $\mu_j(x)\quad\text{and variance}\quad\sigma_j^2(x)$ serve as belief and uncertainty estimates, respectively.

In our implementation, we employ the Mat\'{e}rn kernel~\cite{popovic2020informative2} with separate spatial and temporal length scales. Sensor observations, represented as binary detections indicating the presence or absence of each target within the sensor's field of view, are incorporated as noise-free data points $( (p_\text{obs}, t_\text{obs}), y_\text{obs} )$ to incrementally update the GPs.

\subsection{Environment Representation via Graph Discretization}
To enable efficient planning and computation, we discretize the continuous workspace $\mathcal{W} \subseteq \mathbb{R}^2$ into a graph $\mathcal{G = (V, E)}$, where the node set $\mathcal{V} = \{v_1, ..., v_K\}$ represents monitoring waypoint candidates and the edge set $\mathcal{E}$ defines traversable paths between them. We construct the graph using $k$-nearest neighbor (k-NN) sampling to ensure connectivity, which supports efficient multi-agent policy learning, as agents' observations are tied to their current node locations, their actions correspond to transitions to neighboring nodes, and Gaussian Processes (GPs) inference, which is used for belief updates, is performed only at node positions, significantly reducing computational overhead.

\subsection{Sensing and Constraints}
Each agent has a fixed sensing radius $r_{\mathrm{sense}}$ and can observe the presence of nearby targets within this range. At each time step, for every target \( T_j \), we determine whether it lies within the sensing radius of any agent. If so, the corresponding observation is assigned to the node currently occupied by the observing agent and recorded as a positive binary detection \( y^{(j)} = 1 \); otherwise, \( y^{(j)} = 0 \). These observations are then used to update the Gaussian Process (GP) belief model associated with target \( T_j \), allowing principled belief refinement over time.

To encourage exploration efficiency and enforce practical constraints, we define a global mission budget \( B \) that limits the total number of allowable agent actions throughout the episode.
While bounded, \( B \) is set to be sufficiently large to reflect the persistent nature of the monitoring task.

\begin{figure*}[ht]
\vspace{0.2cm}
\setlength{\belowcaptionskip}{-4mm}
  \centering
    \includegraphics[width=\textwidth]{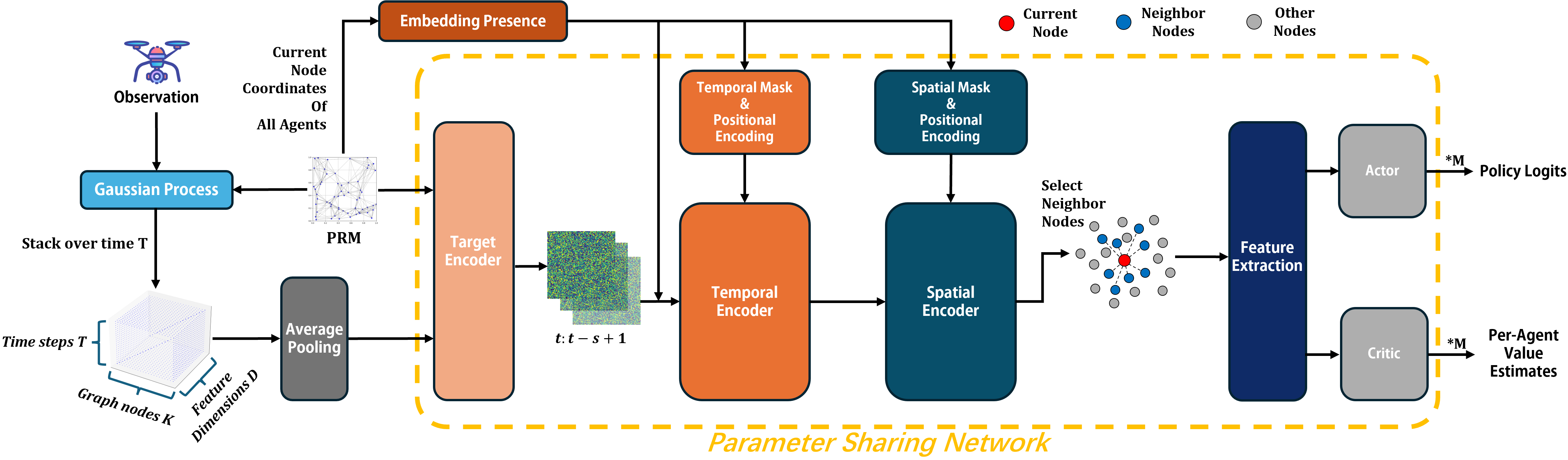}
    \caption{\textbf{Overview of the COMPASS spatio-temporal attention network.} 
    Agent observations are first integrated by Gaussian Processes (GPs) to update belief maps. 
    Encoded historical node features are processed by a \textbf{Temporal Encoder} and a \textbf{Spatial Encoder} with positional and masking mechanisms. 
    The fused representation is used by shared \textbf{Actor} and \textbf{Critic} heads for decentralized policy execution, 
    enabling cooperative uncertainty-aware decision-making among agents.}
    \label{fig:framework}
    \vspace{-0.2cm}
\end{figure*}

\section{Method}
\label{sec:method}

We formulate the multi-agent persistent monitoring task as a Decentralized Partially Observable Markov Decision Process (Dec-POMDP), where a team of agents aims to minimize global uncertainty of target locations through coordinated policies. Each agent observes only local information and must act independently, while the team collectively reduces overall uncertainty through decentralized execution. To solve this, we design a deep reinforcement learning (DRL) framework that integrates Gaussian Process-based belief estimation, spatio-temporal attention-based policy learning, and a graph-based environment representation.

\subsection{Problem Modeling and Belief Estimation}
\label{subsec:method_gp}

We deploy a team of \(M\) agents on the discretized graph $\mathcal{G=(V,E)}$ to monitor \(N\) mobile targets that evolve freely in the continuous workspace \(\mathcal{W}\). By constraining agent motion to graph edges, we significantly reduce planning complexity, while still allowing the targets to move in a continuous spatial field. Time advances in action–observation cycles indexed by \(t=1,2,\dots,T\).

At each decision step, agents rely on a belief representation constructed using Gaussian Processes (GPs), as described in Section~\ref{subsec:gp}. These GPs provide probabilistic estimates of each target’s presence across space and time, and we evaluate their posterior mean and variance at every node for both the current time \(t\) and a short-term future horizon \(t + \delta\). These predictions are concatenated into node-level feature vectors that represent the current and anticipated information landscape. To reflect the deployment state, a binary node presence indicator is appended to each node feature vector, indicating whether it is currently visited by any agent. This representation supports decision-making under uncertainty by informing agents of each other’s locations, enabling coordinated spatio-temporal task allocation via the attention-based policy network. Agents share compact measurement updates during execution: after each measurement cycle, the raw measurements from all agents are aggregated and broadcast globally. Each agent then uses this identical set of observations to independently update its local GP, ensuring the belief maps across the team remain synchronized. Thus, although the policy is executed locally, coordination occurs via this global sharing of raw sensor data.

\subsection{Reinforcement Learning Setup}
\label{subsec:method_rl}

In our formulation, each agent interacts with the environment through local observations and selects navigation actions based on its neighborhood structure. The system evolves over discrete decision steps $k = 0, 1, 2, \dots$, and the learning objective is to maximize cumulative reward over time while guiding agents to explore informative and complementary regions of the environment.

\textbf{Observation Space.}
At each step $k$, agent $m$ observes a local representation $o_m(k)$, which consists of the temporally encoded feature $e_{\text{final}}(v_{\text{loc}(m,k)}, t_k)$ of its current node, along with the feature embeddings of its $K$ neighboring nodes. These features encode GP predictions $(\mu_j, \sigma^2_j)$ for each target $j$, agent presence indicators, 2D spatial coordinates, and node-centric graph position encodings. The resulting observation captures both uncertainty estimates and spatial deployment context, enabling agents to reason over where information gain is likely to be maximized over the entire task horizon.

\textbf{Action Space.}
Given its current location $v_{\text{loc}(m,k)}$, agent $m$ selects an action $a_m(k)$ from its discrete neighborhood set $\mathcal{N}_m(k) \subset \mathcal{V}$, where $\mathcal{N}_m(k)$ contains nodes that are one-hop reachable via graph edge transitions. The state transition for agent $m$ is defined as:
\[
v_{\text{loc}(m,k+1)} = a_m(k), \quad \text{where } a_m(k) \in \mathcal{N}_m(k).
\]
This formulation naturally constrains motion to the physical topology of the sensing environment.

\textbf{Reward Function.}
To encourage efficient and informative exploration, we define a step-wise reward as a weighted combination of three components: information gain, redundancy penalty, and path cost. The reward at decision step $k$ is:
\[
R(k) = \alpha_{\text{info}}(k) \cdot \mathrm{IG}(k)
     - \alpha_{\text{cov}}(k) \cdot \mathrm{CP}(k)
     - \alpha_{\text{path}}(k) \cdot \mathrm{PP}(k),
\]
where $\mathrm{IG}(k)$ quantifies the cumulative information gain across all targets, computed as the total reduction in GP posterior variance, $\mathrm{CP}(k)$ penalizes redundant observations by scaling their contributions based on the remaining uncertainty at each location, thereby discouraging multiple agents from repeatedly visiting the same well-explored nodes, and $\mathrm{PP}(k)$ reflects overall trajectory cost, measured as the sum of shortest-path distances each agent travels between consecutive decision steps on the graph.

The weight coefficients are adaptively scheduled over time via a linear curriculum: $\alpha_{\text{info}}(k) = 3.0 - 1.5 \cdot \rho(k)$, $\alpha_{\text{cov}}(k) = 0.1 + 0.3 \cdot \rho(k)$, and $\alpha_{\text{path}}(k) = 0.05 + 0.05 \cdot \rho(k)$, where $\rho(k) = \min(k / 20000, 1.0)$ denotes normalized training progress. This scheduling strategy emphasizes aggressive information gathering during early training (via a high $\alpha_{\text{info}}$), when uncertainty is widespread, and gradually shifts toward encouraging spatial coverage and movement efficiency. Such a curriculum fosters a natural transition from exploration to refinement, aligning learning incentives with different training stages.

\subsection{Network Architecture}
\label{subsec:method_feature}

Our COMPASS network consists of a shared spatio-temporal attention model used by all agents to extract meaningful features from historical observations and graph structure. The network is composed of a temporal decoder to capture temporal dependencies, a spatial encoder to model inter-node relationships, and policy and value heads to produce action probabilities and value estimates. All agents share parameters of a single network, and experience collected by all agents is aggregated to jointly optimize the network via centralized training.

\paragraph{\textbf{Feature Embedding and Input Projection}} Each node maintains a buffer containing features extracted over the past $H$ timesteps. These include GP-derived mean and variance for each target, binary presence flags indicating agent presence, and spatial coordinates. Average pooling with stride $s$ is applied to compress the temporal dimension, yielding a feature sequence of shape $H' \times d_f$. Each component is embedded into a common feature space through dedicated linear projections. A final projection layer maps the concatenated embeddings to a fixed embedding size $d_e$, resulting in the initial node embeddings.

\paragraph{\textbf{Temporal Encoder}}  
For each node \(v\), we maintain a pooled history \(\mathcal{H}_v = \{\mathbf{h}_{v,t-H+1}, \dots, \mathbf{h}_{v,t}\}\). We treat the most recent vector \(\mathbf{h}_{v,t}\) as the query, and the full history \(\mathcal{H}_v\) as keys and values, \(\mathbf{W}^Q, \mathbf{W}^K, \mathbf{W}^V\) are learnable weight matrices, i.e. \  
\[
\mathbf{Q} = \mathbf{h}_{v,t}\mathbf{W}^Q,\quad \mathbf{K} = \mathcal{H}_v\mathbf{W}^K,\quad \mathbf{V} = \mathcal{H}_v\mathbf{W}^V.
\]  
Multi-head attention then computes a fused embedding  
\begin{equation*}
\begin{split}
\mathbf{e}_{\text{temp}}(v,t) &= \mathrm{Concat}(\mathrm{head}_1,\dots,\mathrm{head}_{N_h})\,\mathbf{W}^O,\\
\mathrm{head}_i &= \mathrm{Softmax}\!\bigl(\tfrac{\mathbf{Q}_i\mathbf{K}_i^\top}{\sqrt{d_k}}\bigr)\,\mathbf{V}_i
\end{split}
\end{equation*}

This temporally-contextual embedding is then forwarded to the spatial encoder.

\begin{table*}[ht]
  \centering
  \caption{
    Performance comparison across different graph sizes ($K \in \{100, 200, 400\}$) and agent team sizes ($M \in \{2, 3, 5\}$). For each metric, $\downarrow$ indicates lower is better, and $\uparrow$ indicates higher is better. Results are averaged over 10 evaluation runs.
  }
  \label{tab:general}
  \resizebox{\textwidth}{!}{%
  \begin{tabular}{@{}ll|ccc|ccc|ccc|ccc|ccc@{}}
    \toprule
    \multicolumn{2}{c|}{\textbf{Metric}} & \multicolumn{3}{c|}{COMPASS} & \multicolumn{3}{c|}{STAMP*} & \multicolumn{3}{c|}{Auction} & \multicolumn{3}{c|}{Coverage} & \multicolumn{3}{c}{Random} \\
    \cmidrule(lr){3-5} \cmidrule(lr){6-8} \cmidrule(lr){9-11} \cmidrule(lr){12-14} \cmidrule(lr){15-17}
     & $K$ & $M{=}2$ & $M{=}3$ & $M{=}5$ & $M{=}2$ & $M{=}3$ & $M{=}5$ & $M{=}2$ & $M{=}3$ & $M{=}5$ & $M{=}2$ & $M{=}3$ & $M{=}5$ & $M{=}2$ & $M{=}3$ & $M{=}5$ \\
    \midrule
    \multirow{3}{*}{Avg Unc $\downarrow$}
    & 100 & \textbf{0.49} & \textbf{0.42} & \textbf{0.37} & 0.50 & 0.53 & 0.48 & 0.66 & 0.63 & 0.59 & 0.76 & 0.70 & 0.63 & 0.80 & 0.78 & 0.71 \\
    & 200 & \textbf{0.52} & \textbf{0.45} & \textbf{0.40} & 0.57 & 0.55 & 0.50 & 0.72 & 0.66 & 0.60 & 0.79 & 0.75 & 0.69 & 0.85 & 0.82 & 0.75 \\
    & 400 & \textbf{0.53} & \textbf{0.46} & \textbf{0.41} & 0.60 & 0.58 & 0.52 & 0.74 & 0.70 & 0.64 & 0.83 & 0.82 & 0.75 & 0.87 & 0.84 & 0.76 \\
    \midrule
    \multirow{3}{*}{Avg JSD $\downarrow$}
    & 100 & \textbf{0.16} & \textbf{0.12} & \textbf{0.10} & 0.17 & 0.18 & 0.14 & 0.27 & 0.22 & 0.19 & 0.37 & 0.32 & 0.27 & 0.43 & 0.41 & 0.35 \\
    & 200 & \textbf{0.18} & \textbf{0.14} & \textbf{0.11} & 0.21 & 0.19 & 0.16 & 0.30 & 0.24 & 0.20 & 0.39 & 0.35 & 0.29 & 0.45 & 0.43 & 0.37 \\
    & 400 & \textbf{0.20} & \textbf{0.17} & \textbf{0.13} & 0.24 & 0.21 & 0.17 & 0.33 & 0.25 & 0.22 & 0.42 & 0.39 & 0.33 & 0.47 & 0.46 & 0.39 \\
    \midrule
    \multirow{3}{*}{Min Visits $\uparrow$}
    & 100 & \textbf{10.1} & \textbf{12.3} & \textbf{18.4} & 9.9 & 8.8 & 10.5 & 4.6 & 6.1 & 7.2 & 2.8 & 3.5 & 5.2 & 3.2 & 4.2 & 5.5 \\
    & 200 & \textbf{9.4} & \textbf{11.8} & \textbf{17.3} & 8.8 & 8.6 & 10.0 & 4.8 & 6.3 & 7.8 & 2.9 & 3.7 & 5.4 & 3.1 & 3.7 & 5.0 \\
    & 400 & \textbf{9.2} & \textbf{11.1} & \textbf{14.2} & 8.4 & 8.6 & 9.7  & 4.5 & 6.3 & 7.5 & 3.0 & 3.7 & 5.5 & 1.7 & 2.5 & 4.2 \\
    \midrule
    \multirow{3}{*}{Avg Visits $\uparrow$}
    & 100 & \textbf{13.5} & \textbf{16.5} & \textbf{23.9} & 13.1 & 11.2 & 13.5 & 6.2 & 7.8 & 9.1 & 4.1 & 5.5 & 7.5 & 4.5 & 6.1 & 7.8 \\
    & 200 & \textbf{13.3} & \textbf{16.2} & \textbf{21.5} & 11.8 & 10.8 & 12.8 & 6.3 & 7.5 & 9.3 & 4.0 & 4.8 & 7.2 & 4.2 & 4.6 & 6.9 \\
    & 400 & \textbf{13.0} & \textbf{15.1} & \textbf{19.0} & 10.9 & 9.2 & 11.5 & 5.8 & 6.9 & 8.4 & 3.4 & 3.5 & 6.5 & 3.2 & 3.2 & 5.8 \\
    \midrule
    \text{Inference Time (ms) $\downarrow$} 
     & 200 & 12.5 & 13.8 & 15.9 
            & 8.2 & 8.5 & 9.0 
            & 2.5 & 2.9 & 3.2 
            & 0.1 & 0.1 & 0.2 
            & 0.1 & 0.1 & 0.1 \\
    \bottomrule
  \end{tabular}%
  }
\end{table*}

\paragraph{\textbf{Spatial Encoder}}
To capture spatial correlations, the temporal embeddings are processed by a Transformer encoder that attends across all nodes. Each node representation is augmented with spatial priors, including a Laplacian positional encoding and a current presence embedding. The resulting spatial embedding is then concatenated with the shortest path distance to the nearest agent and projected back to the embedding dimension, yielding the final node feature $e_{\text{final}}(v, t)$.

\paragraph{\textbf{Policy and Critic Heads}}
For each agent, the policy head queries the features of its $K$ neighboring nodes using a single-head attention mechanism. The resulting attention scores form the policy output $\pi(a|s)$, a probability distribution over the candidate action set. In parallel, a decentralized critic head estimates the state value $V(s)$ from the agent’s current node embedding. Both heads are driven by the same shared backbone, enabling joint optimization and decentralized execution.

\subsection{Training Settings}
\label{subsec:method_training}

We train the shared network using Proximal Policy Optimization (PPO) via centralized training. For data collection, we run $N_{\text{env}} = 16$ parallel environments, gathering trajectories of length $T=100$ from all agents into a shared experience buffer for joint gradient updates.

The policy is optimized using PPO's clipped surrogate objective with a clipping factor of $\epsilon = 0.2$:
\begin{equation}
\begin{aligned}
L^{\text{PPO}} = \mathbb{E}_k \big[ \min(&\ r_k(\theta) \hat{A}(k), \\
                                       &\ \text{clip}(r_k(\theta), 1 - \epsilon, 1 + \epsilon)\hat{A}(k)) \big]
\end{aligned}
\end{equation}
We estimate advantages using GAE ($\gamma = 0.99$, $\lambda = 0.95$) and supplement the loss with a value function term and an entropy bonus. The policy is updated using the Adam optimizer with a learning rate of $10^{-4}$, decayed exponentially. Total training took approximately 20 GPU-hours on two NVIDIA A100 GPUs.

\section{Evaluation}
\label{sec:eval}

\subsection{Experimental Setup}
\label{subsec:eval_setup}

\paragraph{Simulation Environment}
We conduct all experiments in simulated environments. Evaluations including those in Table~\ref{tab:general} and Table~\ref{tab:ablation}, are performed in a scripted simulator with visualization. We also validate our framework in AirSim for high-fidelity simulation, as shown in Fig.~\ref{fig:sim_eval}, which captures realistic spatial dynamics and multi-agent coordination in 3D settings. The workspace $\mathcal{W}$ is a continuous 2D area discretized into a graph $\mathcal{G}$, where $K$ nodes are uniformly sampled in the unit square $[0,1]^2$. Each node is connected to its $k=10$ nearest neighbors to ensure connectivity. Unless otherwise specified, we set $K = 200$, $r_{\mathrm{sense}}=0.1$ and use a default speed factor of $0.6$, indicating that targets move at $60\%$ of the agents’ speed.

The system includes $M=3$ mobile agents tasked with monitoring $N=8$ dynamic targets, each following nontrivial continuous trajectories to simulate realistic spatial drift. Agent sensing is modeled as a circular region with fixed radius $r_{\mathrm{sense}}$, and the belief about each target's distribution is maintained via Gaussian Processes (GPs), which are updated upon successful observation. Each evaluation episode is constrained by a mission budget $B=30$, denoting the maximum allowed action steps per agent.

\paragraph{Evaluation Metrics}
To assess the performance of each method, we adopt four quantitative metrics. (1) \textbf{Average Uncertainty (Avg Unc)} measures the mean posterior standard deviation aggregated over all targets, spatial nodes, and time steps; it captures the overall confidence in the learned belief, with lower values indicating more accurate and reliable estimates. (2) \textbf{Average Jensen-Shannon Divergence (Avg JSD)} quantifies the average divergence between the GP-estimated belief distribution and the ground truth distribution of target locations, computed over time and space; lower values indicate better alignment and belief fidelity. (3) \textbf{Minimum Target Visits (Min Visits)} records the smallest number of successful observations made for any target in an episode; it reflects worst-case neglect and the ability to maintain balanced monitoring across all targets, with higher values being preferred. (4) \textbf{Average Target Visits (Avg Visits)} captures the average number of observations across all targets during an episode, representing overall sensing effort and temporal coverage density; again, higher values are better.

\paragraph{Baseline Methods}
We compare our proposed method, COMPASS, with four baselines:

\begin{itemize}
    \item \textbf{STAMP*}: A decentralized adaptation of the STAMP framework ~\cite{stamp}, where each agent independently runs its own policy and value networks without communication or parameter sharing. Each agent relies solely on local observations and historical context to determine its action sequence.
    
    \item \textbf{Auction}: A distributed coordination strategy where agents compete for targets using an auction-style bidding mechanism. Each agent evaluates potential targets based on their spatial proximity and associated uncertainty and selects actions that maximize its local utility score. No centralized planner or global policy is used.
    
    \item \textbf{Coverage (Lawnmower)}: A non-reactive strategy in which agents follow predefined paths to systematically traverse the environment. Paths are initialized from a single global route (e.g., TSP) and divided among agents. This strategy ignores current uncertainty or target locations, aiming only for spatial coverage.
    
    \item \textbf{Random}: At each time step, agents randomly select a neighbor node to move to, without regard for any belief state or historical observation. This baseline represents an uninformed exploratory behavior and serves as a lower-bound reference.
\end{itemize}

\subsection{Performance Analysis and Scalability}
\label{subsec:eval_scalability}

Table~\ref{tab:general} reports comparative results across varying graph sizes ($K$) and team sizes ($M$). COMPASS consistently achieves the lowest Average Uncertainty (Avg Unc) and Jensen-Shannon Divergence (Avg JSD), indicating superior belief estimation and uncertainty reduction. As $K$ increases, all methods experience moderately higher uncertainty due to expanded spatial areas and reduced node density. However, COMPASS exhibits minimal degradation (e.g., Avg Unc rises only from 0.42 to 0.46 for $M=3$), while baselines like STAMP* and Auction deteriorate more noticeably. It also maintains belief fidelity at small $M$ (e.g., $M=2$) and benefits more from larger teams, underscoring its ability to coordinate agents effectively in sparse environments.

Coverage-related metrics reinforce this trend. COMPASS achieves the highest Minimum and Average Target Visits across all settings, reflecting strong cooperation and efficient budget use. The gains with increasing $M$ are particularly notable (e.g., Avg Visits rises from 16.5 to 23.9 as $M$ grows from 3 to 5 under $K=100$), while baselines either saturate or improve marginally. Even under more challenging $K=400$ scenarios, COMPASS sustains better coverage and revisit balance. These results confirm that COMPASS scales robustly with environmental complexity and team size, making it a strong candidate for real-world persistent surveillance.

\begin{figure}[htbp]
  \centering
  \includegraphics[width=0.45\textwidth]{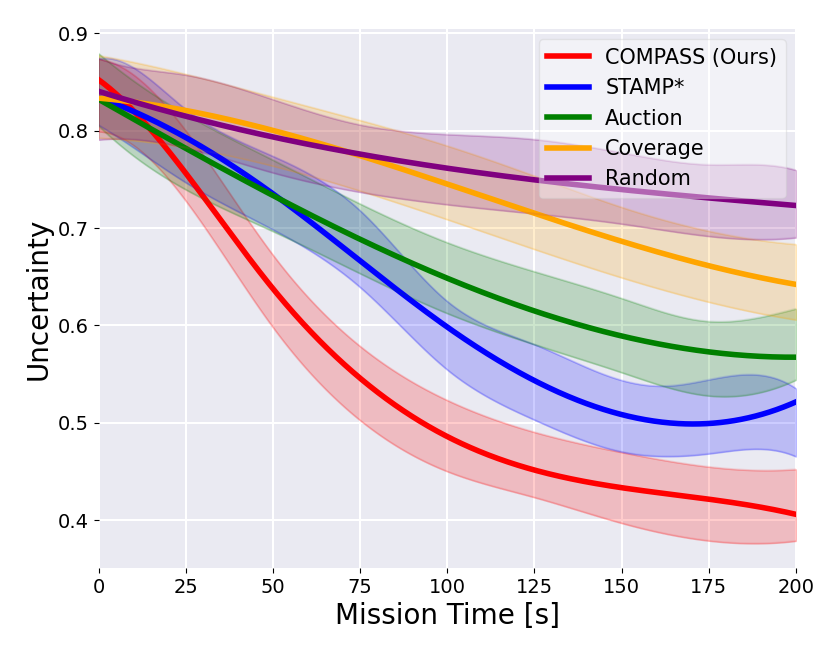} 
  \caption{\textbf{Average uncertainty over mission time.} 
  COMPASS achieves the fastest and most stable uncertainty reduction compared to baseline methods ($K=200, M=3, N=8$). 
  Solid lines show means; shaded areas denote standard deviations over 20 runs.}
  \label{fig:sim_eval}
  \vspace{-0.2 cm} 
\end{figure}

Figure~\ref{fig:sim_eval} shows the temporal evolution of average uncertainty under the default configuration ($K = 200$) in simulation. COMPASS rapidly reduces uncertainty during the early phase and converges to the lowest level, sustaining it through the mission. STAMP*, lacking centralized value estimation, reduces uncertainty slower and plateaus at a higher level. Auction exhibits fast initial drops but stagnates due to its myopic nature. Coverage and Random perform poorly throughout. These trends confirm that COMPASS enables fast, consistent, and well-coordinated information gathering over time.

\subsection{Ablation Study}
\label{subsec:ablation}

To assess the contribution of each component in COMPASS, we conduct an ablation study under the default setting ($K = 200$, $M = 3$, $N = 8$, $B = 30$). We report average values across evaluation episodes for three key metrics: \emph{Uncertainty}, \emph{RMSE}, and \emph{Target Visits}. Specifically, we evaluate the following three ablated variants:
(1) \textbf{No PresenceInfo}: Removes the presence indicator from node features, preventing agents from observing each other’s current positions on the graph.
(2) \textbf{No SpatialAttn}: Removes the spatial attention module and replaces it with mean pooling over each node’s spatial neighbors.
(3) \textbf{No TemporalAttn}: Removes the temporal attention module and applies uniform averaging over historical node embeddings.

Table~\ref{tab:ablation} shows that removing spatial attention leads to the largest degradation, with Uncertainty and RMSE both increasing, and fewer targets being visited. This highlights the importance of learning spatial dependencies to guide agents toward informative regions. The temporal attention module also plays a key role; without it, the agent's ability to track dynamic targets is impaired, resulting in reduced visit counts and degraded accuracy. Removing presence encoding leads to a moderate drop, suggesting that while spatial context is beneficial, attention mechanisms contribute more significantly to model performance.

\begin{table}[t]
  \centering
  \caption{Ablation results under default configuration ($K = 200$). All values are averaged over 10 runs. Lower Uncertainty and RMSE are better; higher Visits is better.}
  \label{tab:ablation}
  \resizebox{0.8\columnwidth}{!}{%
  \begin{tabular}{@{}lccc@{}}
    \toprule
    \textbf{Variant} & \textbf{Uncertainty} & \textbf{RMSE} & \textbf{Visits} \\
    \midrule
    \textbf{Full COMPASS}    & \textbf{0.45} & \textbf{0.26} & \textbf{16.5} \\
    No PresenceInfo             & 0.53 & 0.35 & 14.6 \\
    No SpatialAttn           & 0.57 & 0.35 & 12.1 \\
    No TemporalAttn          & 0.56 & 0.29 & 13.7 \\
    \bottomrule
  \end{tabular}
  }
  \vspace{-0.4 cm}
\end{table}

\section{Conclusion and Future Work}
\label{sec:conclusion}

This paper introduced COMPASS, a decentralised framework that couples GP-based belief estimation with a shared spatio-temporal attention backbone.  Agents trained with a central-critic PPO rely only on local observations yet coordinate implicitly, and experiments show up to \(20\%\) lower mean uncertainty together with a twofold improvement in worst-case visitation frequency relative to strong learning and heuristic baselines.  Ablation studies verify that spatial and temporal attention are both indispensable, with spatial reasoning providing the greater share of the gain.

Although our simulator already includes stochastic target motion and limited sensing, it still omits many real-world complexities.  Future work will narrow this gap through domain randomisation, robust policy training, and model-adaptation techniques that compensate for sensor noise, localisation drift, occlusion, and actuation limits.  Using high-level waypoint control rather than low-level motor commands is a deliberate design choice that simplifies transfer across robots and promotes scalability, yet it may sacrifice fine-grained agility; exploring hybrid control stacks that blend the two levels is therefore an interesting direction.  We also plan to enrich cooperation via explicit yet bandwidth-aware communication and to validate COMPASS on physical platforms operating over extended horizons.  By providing an uncertainty-aware and computationally efficient blueprint, COMPASS moves a step closer to field-ready autonomous surveillance systems.

\bibliographystyle{unsrt}
\bibliography{bib}

\addtolength{\textheight}{-12cm}   





\end{document}